# A possible universal definition for the nanophase

M Ghanashyam Krishna, M Durga Prasad[#], V Srinivasan[*]

School of Physics, University of Hyderabad, Hyderabad-500 046, India

[#]School of Chemistry, University of Hyderabad, Hyderabad-500 046, India

Abstract

A possible universal definition for a nanostructured material based purely on experimental data available in literature for a wide variety of physical phenomena is proposed. It is suggested that for values of the ratio of sample volume, V to that of the unit cell volume, $V_c$, $\leq 10^5$- $10^6$ the samples behave as nanostructures. This is mirrored in the ratio of the number of particles in sample, N to that in the unit cell, $N_c$. It is further proposed that the transition to the nanophase from bulk is a phase transition in all cases investigated. The nanophase should therefore be treated as a distinct phase of matter, entirely in the quantum mechanical domain, and treated appropriately.

*Corresponding Author: EMAIL: vssp@uohyd.ernet.in

Phone: (+)91-40-23134351

FAX: (+)91-40-23010227

Over the last decade or so the preparation of nano-structured materials, theoretical investigations thereof and applications of these have evoked wide interest. Intuitively designed nano-materials based on the currently available tools have been realised and their physico-chemical properties investigated [1]. Intuition demands that the volume of a nano-material should be small. However, the limitations on the size of sample to be defined as a nano-material are not apriori known; nor do universal theoretical calculations exist that quantify this phenomenon. Phenomena based model calculations are available in some cases, however, a universal quantified definition of the nanophase is still not forthcoming. In this paper, we address this from a phenomenological point of view and base our conclusions strictly on the large body of experimental evidence available in literature.

The first underlying assumption made is that the laws of quantum mechanics strictly govern the nanophase, and therefore the phenomena observed should be explainable only from quantum mechanics. This should be contrasted with quantum field theory effects such as superconductivity, ferromagnetism and superfluidity that are ground state phenomena obeyed in the bulk. In other words, we strictly restrict ourselves to the quantum mechanical domain without invoking the infinite volume limit. Since size is an important factor, the ratio of surface area to volume is large for small systems. i.e.

$(V^{2/3}/V) = 1/V^{1/3}$

becomes smaller and smaller as V grows. The surface effects, therefore, dominate in the nanophase in contrast to the bulk state. So the question that emerges is how small should the system be to be defined as a nano-phased material. A similar question arises in crystallography. A phenomenological approach considered by us recently [2], came to the conclusion that for $V/V_c < 10^6$ the transition to the nanophase is made where V is the volume of crystallites in the solid and $V_c$ the unit cell volume. A similar analysis will be used here. In fact all second order transitions occur only when N and V both tend to infinity but N/V is finite. However, this is only a theoretical limit. The approach used here to define the limiting volume of the nanophase interestingly answers the other question, i.e. the volume tending to infinity limit. This seems to apply even to non-equillibrium phase transitions such as lasers.

The nanophase of a material need not be unique. Several possible structures allowed by quantum mechanics are possible. Furthermore, one can go to the nanophase from solid, liquid or gas. It is clear that a criterion for the transition can only emerge from experiments that are able to cause a gradual decrease from bulk-like sizes until a critical size is reached beyond which behaviour is completely governed by quantum mechanics. We shall analyze this by examining a wide variety of physical phenomena where the volume is decreased slowly and nanostructures emerge at some point. If the phenomenon is observed in solids for which the volume of the unit cell is known, it can be used as a natural unit of volume. In the case of liquids and gases the coefficient 'b' in the Van der Waals equation is an experimentally verified quantity and is therefore the analogue of the volume of the unit cell. Indeed, the order of magnitude of this is the same as that of the unit cell. Examples of transitions to the nanophase of a wide variety of physical phenomena are shown in Table I.

From these wide variety of examples it emerges that if the experimentally demonstrated sample size is used as the limit then nano-phase is observed upto the upper bound of $V/V_c < 10^6$. This volume ratio can be converted in to a ratio for the number of particles by substituting the number of atoms per unit cell. In Bravais lattices, it is well-known that this number rarely exceeds four. Therefore it leads to the ratio of $N/N_c < 10^4$. Indeed this seems to be obeyed by Bose-Einstein condensates. The condensates occur only when $N \geq 10^4$ . This is also true for lasers where population inversion occurs only when $N>10^7$ for a He-Ne laser. This number is slightly higher due to the dissipation effects in the cavity. It should be noted that no particular solutions of Euler equations or their like should be found in nanostructures. These correspond to the ground states of bulk systems and therefore vortices, dislocations, defects and Bloch walls etc. cannot occur in nanostructures. This is in accordance with experiments wherein nanostructures have been shown to exhibit greater hardness than their bulk forms.

The continuum limit, where the transition is from lattice quantisation to field theory, relies on the modes whose frequencies go to a continuous set of frequencies. In practice no system is infinite and therefore it is difficult to evolve a mathematical criterion that can set the limit of transition from finite to infinite volume based on a physical parameter. With the emergence of nanomaterials, this question becomes more important since non-relativistic quantum mechanics is operable in this regime while field theory and the grand canonical ensembles are operable only in the V and N tending to infinity limit. The $N/N_c$ ratio can be taken as the ratio that defines the transition to bulk behaviour such that for $N/N_c > 10^5$ the solid behaves as bulk.

Some examples where the bulk to nanophase transition can be seen are presented in Table I. For instance, one of the oldest known examples of such a transition is the size dependent ferromagnetic to superparamagnetic transition [3]. The superparamagnetic behaviour is completely quantum mechanical in nature while classical paramagnetism is not. In Bose-Einstein condensation, only when the number of particles in the cloud is $10^5$ is the superfluid behaviour observed [4]. At this value of the ratio, all examples shown in the table including these two show a transition to nanostructured behaviour. The bulk to nanophase transition can occur in solid, liquid or gaseous state as permitted by the Van der Waal's equation (for the bulk state). The saturated vapour pressure in equillibrium with droplets of liquids is higher compared to the saturated vapour pressure over the bulk liquid [5]. This relation given by the Kelvin equation predicts that the droplet size is 40 nm for a difference of 1% in the two vapour pressures for a surface tension of 20 mN $m^{-1}$. This size, significantly, represents about $10^5$ molecules. Recently $^4$He clusters in the liquid state were obtained [6]. On laser cooling these, the nanostructure to bulk transition occurs when the clusters go to the superfluid state, whereas the clusters are gaseous at normal temperatures. These examples suggest that there is a phase transition when $V/V_c$ is of the order of $10^6$. This is even more evident from figures 1 and 2 where the data from some of the examples presented in table 1 have been replotted as a function of $V/V_c$. These figures show that the collapse of superconducting transiton in YBCO (fig.1), appearance of superparamagnetism in Cobalt ferrite (fig. 1), structural phase transition in Co particles (fig. 2) and ferroelectric to paraelectric transition in $BaTiO_3$ (fig. 2) all occur when $V/V_c$ is of the order of $10^6$. Furthermore, the behaviour indicates clearly the occurrence of a phase transition at this limiting value.

Since all transitions go to a different phase in their physical characteristic described by a different phenomenon from that determining bulk behaviour, we propose that the nano-bulk transition is a phase transition. Also, recent experiments suggest that the nanophase, as it should, is retained as temperature is raised while the bulk undergoes a phase transition from solid to liquid, the nanophase does not do so [14]. Experiments performed on nano-water show that from $10^{o}$K to $300^{o}$K the large mean square displacement of hydrogen, the physical parameter used by them, does not undergo any abrupt changes, while in the bulk, ice transforms in to water and this physical parameter changes abruptly. Recent experiments on carbon nanotubes show that the additional Raman lines, that are not present in the bulk phase, appear in the nanophase. To a first approximation it seems that the $V/V_c$ even in this case is $10^6$ [15]. This further confirms our conclusion that nano is indeed a new phase, distinct from the solid, liquid and gas phases that appear in the bulk. If $V/V_c$ is greater than $10^5$-$10^6$ depending on the temperature it will go to one of the phases. The chemical potential, $\mu$ , is zero for the nanophase while it is non-zero for the bulk form. The Gibbs free will therefore abruptly change when $V/V_c$ exceeds the limit of $10^6$. This seems to be the limit above, which the system should be, considered a Grand Canonical Ensemble. The Gibbs free energy in general is a function of Temperature (T), Pressure (P) and average number of particles (N). The chemical potential, $\mu$, per particle is non-zero for a grand canonical ensemble. For a nanostructure which is governed by quantum mechanics only and has a finite number of particles, $\mu$ is zero. Therefore, there is an abrupt change in the Gibbs function as a function of N when the limit $N_c$ is exceeded. These arguments originate from general considerations of thermodynamics. Calculations, in particular cases, will only give the numerical values for that system.

The Gibbs free energy is

$$G = \theta( N_c - N) G_{nano} + \theta(N-N_c)G_{bulk}$$

In order to fit into Gibbs's classification the θ function can be replaced by the equivalent smooth function. Since there is an abrupt change in the Gibbs free energy, this is a new kind of phase transition that does not fit into Gibbs classification.

In short we propose that besides solid, liquid and gaseous states the nanophase should be considered as a distinct state of matter and treated appropriately. The signal for the onset of the nanophase is the phase transition.

**Acknowledgements**

The authors wish to acknowledge several discussions with our dear friend late Professor Bhaskar Maiya, who passed away very unexpectedly in March, 2004. Numerous discussions and suggestions of Professor Subhash Chaturvedi are very gratefully acknowledged.

Table I: Examples of bulk-nanophase transitions in different physical phenomena

| Phenomenon | Bulk state phase | Nano-state | $V/V_c$ | Reference |
|---|---|---|---|---|
| Magnetism | Ferromagnetic (solid) | Superparamagnetic (solid) | $10^5$ | 3 |
| Superfluidity | Bulk $^4$He (liquid) | $^4$He Clusters | $10^5$ | 4 |
| Bose-Einstein Condensate | $^{40}$K Bose Superfluid | | $10^5$ | 6 |
| Superconductivity | Bulk Superconductor (solid) | No Superconductivity (solid) | $10^6$ | 7 |
| Optical absorption | Solid with single crystal like band gap | Solid with blue shifted band gap (qunatum confinement) | $10^5$ | 8 |
| Structural phase transition | Co (hcp) | Co(fcc) | $10^6$ | 9 |
| Structural Phase transition | Fe-Ge (fcc) | bc | $10^6$ | 10 |
| Superplasticity | Cu (normal metal) | Cu (superplastic solid) | $10^6$ | 11 |
| Multi-domain to single domain | $La_{0.875}Sr_{0.125}MnO_3$ (Multidomain solid) | $La_{0.875}Sr_{0.125}MnO_3$ (single domain solid) | $10^6$ | 12 |
| Ferroelectricity | Ferroelectric | paraelectric | $10^6$ | 13 |
| Encapsulated water | Bulk | Nanotube water | | 14 |

**Figure captions**

Figure 1. The variation in $T_c$ of $BaTiO_3$ (dark squares) and the coercivity (open circles) as a function of $V/V_c$ clearly shows the phase transition at $10^6$ in both cases. The figure has been replotted using data from refs. 13 and 3 respectively.

Figure 2. The variation in $T_c$ of YBCO (dark squares) and structural (circles)phase transition in Co particles as a function of $V/V_c$ . The figure has been replotted using data from refs. 7 and 9 respectively.

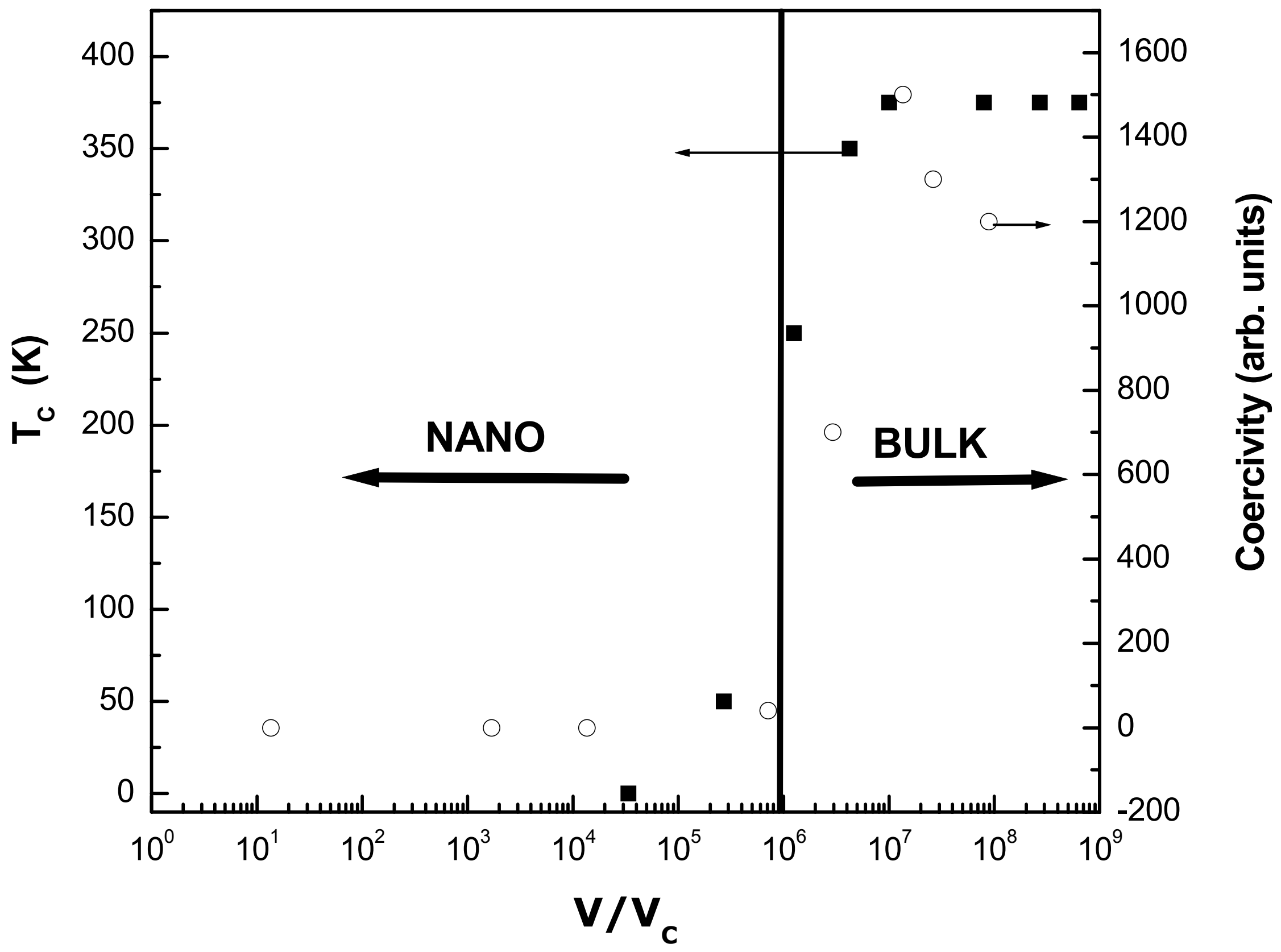

400
350
300
250
200
150
100
50
0
$T_c$ (K)
1600
1400
1200
1000
800
600
400
200
0
-200
Coercivity (arb. units)
NANO
BULK
$10^0$ $10^1$ $10^2$ $10^3$ $10^4$ $10^5$ $10^6$ $10^7$ $10^8$ $10^9$
$V/V_c$

Figure 1

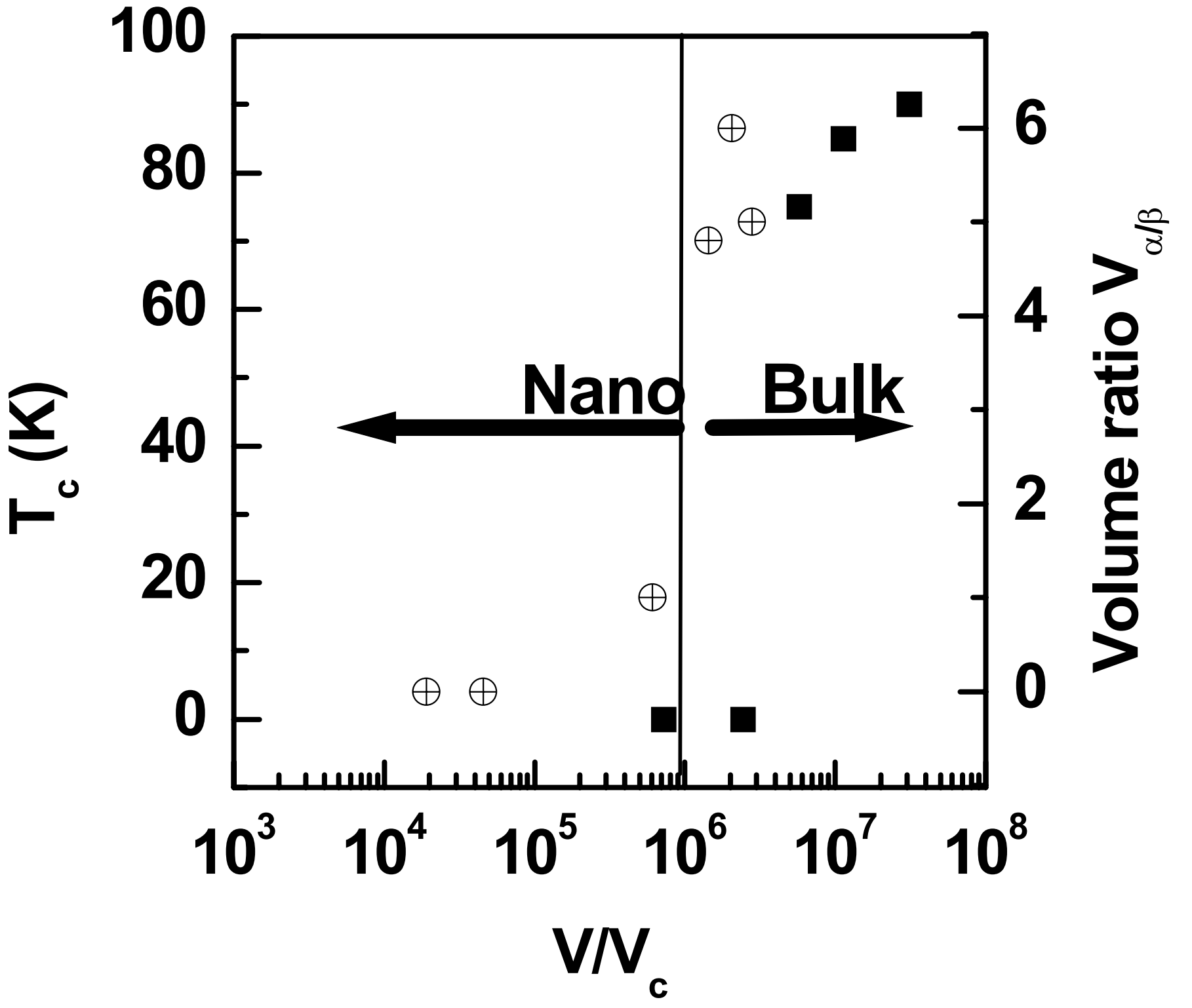

Figure 2